\documentclass[amsmath,jcp,reprint,twocolumn,superscriptaddress]{revtex4-1}

\usepackage{amssymb}
\usepackage{simplewick}
\usepackage{graphicx}
\usepackage{color}

\begin{document}

\title{Nonequilibrium diagrammatic technique for Hubbard Green functions}

\author{Feng Chen}
\affiliation{Department of Physics, University of California, San Diego, La Jolla, CA 92093, USA}
\author{Maicol A. Ochoa}
\altaffiliation{Present Address: Department of Chemistry, University of Pennsylvania, Philadelphia, PA 19104, USA}
\affiliation{Department of Chemistry and Biochemistry, University of California, San Diego, La Jolla, CA 92093, USA}
\author{Michael Galperin}
\affiliation{Department of Chemistry and Biochemistry, University of California, San Diego, La Jolla, CA 92093, USA}

\begin{abstract}
We introduce diagrammatic technique for Hubbard nonequilibrium Green functions (NEGF).
The formulation is an extension of equilibrium considerations for strongly correlated
lattice models to description of current carrying molecular junctions. 
Within the technique intra-system interactions are taken into account exactly, 
while molecular coupling to contacts is used as a small parameter in perturbative expansion. 
We demonstrate the viability of the approach with numerical simulations for a generic junction model
of quantum dot coupled to two electron reservoirs.
\end{abstract}

\maketitle


\section{Introduction}\label{intro}
Using single molecules as electronic devices was originally suggested in a seminal paper
by Aviram and Ratner~\cite{AviramRatnerCPL74}. 
First experimental realization of a single molecule conductance junction was reported 
twenty years later~\cite{ReedTourScience97}. Since then the field of molecular
electronics experienced rapid development~\cite{RatnerNatNano13}; today
molecular junctions are utilized as both nanoscale devices and convenient platforms for
study of fundamental physical properties of matter at nanoscale~\cite{visionsNatNano13,VenkataramanNatNano13}.
Experimental advancements in  measurements of junction responses 
(charge, spin, and energy fluxes; optical scattering; thermoelectric characteristics) 
to external perturbations (bias and gate voltages, temperature gradient, optical and voltage pulses) 
led to a surge in development of corresponding theoretical tools.

Molecular electronics is an interdisciplinary area of research combining condensed matter physics, 
statistical mechanics, nanoplasmonics, nonlinear optics, quantum chemistry, and engineering.
Traditionally theoretical considerations of molecular junctions employ tools 
originally developed in high energy physics and successfully utilized in mesoscopic physics
for description of transport in quantum dot junctions. In particular, the nonequilibrium
Green function (NEGF) technique~\cite{KadanoffBaym_1962,KeldyshSovPhysJETP65,Danielewicz1984,RammerSmithRMP86} 
is the usual choice in {\em ab initio} simulations. 
The quasiparticle (orbital or elementary excitations) basis
employed in NEGF has many important advantages, from the ability to easily treat
systems of realistic sizes to developed diagrammatic perturbation theory which allows
to account for intra-molecular interactions within a controlled expansion in a small interaction
parameter. Density functional theory (DFT)~\cite{ParrYang_1989,DreizlerGross_1990} 
- also formulated in the basis of effective single particle orbitals - 
yields molecular electronic structure. The natural combination
of the two approaches, the NEGF-DFT, was formulated~\cite{DamleGhoshDattaCP02,XueDattaRatnerCP02,TaylorStokbroPRB02} and successfully
applied in many simulations where intra-molecular interaction (e.g., electron-phonon coupling
in the off-resonant tunneling regime) is the smallest energy scale~\cite{GuoPRL05,JauhoPRB07,Schmaus2011,AvrillerFrederiksenPRB12}.

As in any technique, the NEGF-DFT has its limitations; in particular, in treating strong intra-molecular 
interactions. The limitations are especially evident in the resonant tunneling regime 
(regime where the electron tunnels into one of the molecular orbitals), 
which is of particular importance for practical applications~\cite{visionsNatNano13}:
a large response of the molecular structure to external perturbation 
(e.g., negative differential resistance~\cite{ReedRawlettTourScience99,TormaACSNano10,BlumNatMat05,KiehlApplPhysLett06,TourRielSmall06,WeiOrgElectron07,HoJPCC08},
current induced chemistry~\cite{HoJCP02,SeidemanJPCM03}, or switching~\cite{BlumNatMat05,VanDerMollenLiljerothJPCM10,ZantSmall10,KomedaLorenteNatCommun11,CuevasScheerNatNano13})
is a requirement for constructing an effective molecular device.
For example, even such a simple resonant tunneling phenomenon as Coulomb blockade
requires special treatment by DFT; only recently developed hybrid functionals~\cite{BaerNeuhauserPRL05,BaerPCCP07,BaerKronikPRL12,NeatonNL14}
are capable of appropriate description of the effect. Note that the often utilized 
Landauer-DFT version of the NEGF-DFT may even lead to qualitative failures~\cite{BaratzMGBaerJPCC13,BaratzWhiteMGBaerJPCL14}.
We note in passing that NEGF-DFT (and NEGF-TDDFT) approaches have 
also fundamental limitations related to both foundations of (TD)DFT
(utilization of Kohn-Sham orbitals as physical objects, non-uniqueness of the excited-state potentials,
question of stability, and chaos of the mapping of densities on potentials, etc.)~\cite{BurkePRL04,BaerJCP08}
and in combination of DFT with NEGF for transport description~\cite{BaldeaBeilstJNanotech16}.

\begin{figure}[b]
\centering\includegraphics[width=\linewidth]{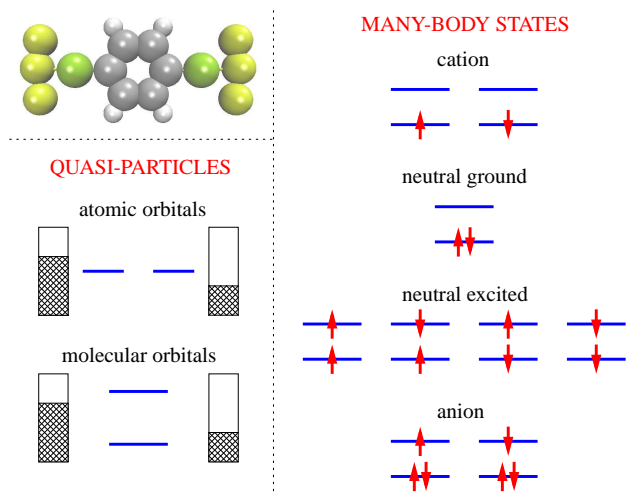}
\caption{\label{fig1}
Sketch of a molecular junction. 
Nonequilibrium atomic limit utilizes basis of many-body states of an isolated molecule
(thus accounting for intra-molecular interactions) while treating coupling between molecule 
and contacts within perturbation theory.
}
\end{figure}

The difficulties mentioned above with NEGF-DFT led to the development of a
nonequilibrium theory in which the system's response to external perturbations is described
using the many-body eigenstates of the isolated molecule (see Fig.~\ref{fig1}).
Such basis may be advantageous in theoretical description of a number of
experiments~\cite{McEuenNature00,WurtzNL07,ReppNatPhys10,NatelsonPCCP13,LiljerothNatPhys15,TautzPRL15}. It also facilitates the incorporation of the methods of quantum chemistry~\cite{YeganehNL09} 
and nonlinear optical spectroscopy~\cite{GaoMGJCP16_2} (usually applied to isolated molecules)
into description of open nonequilibrium molecular systems.
Contrary to the NEGF, where quadratic system-bath coupling is taken into account
exactly while intra-system interactions are treated by perturbation theory,
techniques utilizing many-body states of the system account for the intra-system interactions
exactly while coupling to contacts is treated perturbatively. Such techniques are mostly applicable
for molecules relatively weakly coupled to substrate(s). Similar to the equilibrium
analog we coined this as the nonequilibrium atomic limit~\cite{WhiteOchoaMGJPCC14}.
We note that consequences of treating molecule and its coupling to electrodes at different level
of theory were discussed in the literature~\cite{BaldeaJCP10}.
Validation of such an approach and dependence of results on different partitioning schemes
in realistic simulations is postponed for future research. 

There are two flavors of Green function techniques capable of describing 
response of a nonequilibrium system coupled to environment (e.g., molecule coupled to contacts) 
utilizing the system's many-body states: the pseudoparticle (PP) NEGF and the Hubbard NEGF.
A simplified variant of the former - the slave boson technique - has been utilized to describe 
transport in junctions in seminal papers by Wingreen and Meir~\cite{WingreenMeirPRB94}.
Recent development of the dynamical mean field theory~\cite{WernerRMP14} 
renewed interest in the PP-NEGF~\cite{EcksteinWernerPRB10,OhAhnBubanjaPRB11}. 
The methodology in its lowest (second) order in the system-bath coupling 
(the non-crossing approximation, NCA) was recently applied to describe transport
and optical response in molecular junctions~\cite{OhAhnBubanjaPRB11,WhiteGalperinPCCP12,WhiteFainbergMGJPCL12,WhitePeskinMGPRB13,WhiteTretiakNL14,CohenMillisReichmanPRB16}.
Pseudoparticle Green functions (two-time correlation functions of creation and annihilation
operators of many-body states of the system) can be viewed as a generalization
of density matrix (which describes time-local correlation of many-body states).
The methodology utilizes second quantization in the state space. As a result its formal structure
is completely equivalent to the standard NEGF. However such formulation
is possible only in an extended Hilbert space, whose physical subspace is defined
by the normalization condition (sum of probabilities for the system to be in any of its
states should be one)~\cite{GaoMGJCP16_2}. The latter restriction is the main inconvenience of the method.

The Hubbard NEGF is formulated solely in physical Hilbert space.
It utilizes Hubbard (or projection) operators
\begin{equation}
\hat X_{S_1S_2}\equiv\lvert S_1\rangle\langle S_2\rvert
\end{equation}
on the states $\lvert S_{1,2}\rangle$ of the system. 
Below we distinguish between diagonal (e.g., pair of states containing same number of electrons) 
and non-diagonal operators. For the latter we introduce creation (e.g., number of electrons 
in state $\lvert S_1\rangle$ is bigger compared with $\lvert S_2\rangle$) and 
annihilation operators of Fermi (e.g., difference in number of electrons between the two states
is odd) or Bose type.
Central object of interest is the Hubbard Green function 
(two-time correlation functions of the Hubbard operators)
\begin{equation}
 \label{HubGF}
 G_{(S_1S_2),(S_3S_4)} (\tau,\tau') \equiv -i\langle T_c\,\hat X_{S_1S_2}(\tau)\,
 \hat X_{S_3S_4}^\dagger(\tau')\rangle
 \end{equation}
 Here $\tau$ and $\tau'$ are the Keldysh contour variables, $T_c$ is the contour ordering operator,
and $\hat X_{S_1S_2}$ and $\hat X_{S_3S_4}$ are annihilation operators
in the Heisenberg representation.
Green functions (\ref{HubGF}) are most closely related to the usual NEGF. 
Indeed, while NEGF deals with correlations of elementary excitations (quasiparticles) $\hat c_i$,
\begin{equation}
 \label{GFNEGF}
 G_{ij}(\tau,\tau')=-i\langle T_c\,\hat c_i(\tau)\,\hat c_j^\dagger(\tau')\rangle,
\end{equation}
the Hubbard NEGF yields spectral decomposition of the excitations into
underlying transitions between many-body states of the system,
$\hat c_i=\sum_{S_1,S_2}\langle S_1\rvert \hat c_i\lvert S_2\rangle\,\hat X_{S_1S_2}$,
and considers correlations between pairs of such transitions. 
Thus, knowing the Hubbard NEGFs one always can reconstruct the NEGF. 
Introducing auxiliary fields, one can write the exact equation of motion for 
the Hubbard NEGF (\ref{HubGF}) in terms of the fields and their functional derivatives 
(see, e.g., Ref.~\cite{MGNitzanRatnerPRB08} for details), and solve them approximately. 
For example, ignoring the auxiliary fields in the EOM corresponds
to the first Hubbard approximation (HIA)~\cite{Fransson_2010}. 
The method was applied to transport problems in a number of publications~\cite{FranssonPRB02,FranssonPhotonNanostruct04,FranssonPRB05,SandalovJPCM06,SandalovPRB07,MGNitzanRatnerPRB08,YeganehNL09}.

While applicable to Green functions, the auxiliary field approach is less convenient
in derivations of multi-time correlation functions. Such functions yield information
on optical response~\cite{Mukamel_1995} to classical radiation fields~\cite{GaoMGJCP16_1}, fluctuation theorems, and counting statistics~\cite{EspositoRMP09} of the system.
A possible alternative is diagrammatic perturbation theory, which allows building
controlled approximations for both Green (two-time) and multi-time correlation functions.
Here we  present a nonequilibrium diagrammatic technique for the Hubbard NEGF,
and illustrate its viability within generic model simulations.
The approach is an extension of equilibrium diagrammatic technique for Hubbard operators
in lattice models~\cite{IzyumovSkryabin_1988,OvchinnikovValkov_2004} 
applied to nonequilibrium realm of molecular junctions.
It can be considered as a Green function generalization of the real-time perturbation
theory developed for density matrices~\cite{SchonPRB94,SchoellerSchonPRB96,Schoeller2000,LeijnseWegewijsPRB08}.
As such it illustrates connection between Green function and density matrix methodologies.
We note in passing that density matrix oriented formulations capable to provide efficient
numerical solutions for time-local quantities are available in the literature.~\cite{YanJCP08,MillisPRB13,YanJCP14,YanWIRE16,YanFrontPhys16}

In summary, presented formulation is first nonequilibrium diagrammatic technique
applicable to multi-time correlation functions of Hubbard operators. 
Contrary to standard diagrammatic techniques it utilizes system-bath coupling as 
a small parameter of expansion with intra-system interactions taken into account exactly.
At the same time (similar to standard diagrammatic techniques) diagrams can be 
analyzed and particular subsets responsible for a physical process of interest retained.
Compared to PP-NEGF (similar perturbative expansion) the current technique
is formulated solely in the physical Hilbert space, which makes it potentially
advantageous in formulation of, e.g., the full counting statistics for interacting systems.
Finally, being much lighter than numerically exact methods the approach can be utilized
in first principles simulations of realistic systems.

The structure of the paper is the following: Section~\ref{diag} introduces general diagrammatic 
rules for the Hubbard NEGF. We specialize to generic model of a quantum dot junction 
in Section~\ref{qdmodel}. Section~\ref{numres} presents numerical simulations;
where possible (in noninteracting case) we compare the perturbation theory to exact results. 
Section~\ref{conclude} concludes.


\section{Diagrammatic technique for Hubbard NEGF}\label{diag}
Diagrammatic technique for NEGF is based on the Wick's theorem~\cite{FetterWalecka_1971} 
which relies on (anti)commutation relations for creation and annihilation (Fermi) Bose operators,
$[\hat c_{i};\hat c^\dagger_{j}]_\pm=\delta_{i,j}$. Similarly one has to consider (anti)commutations
of Hubbard operators for a transition $m=(S_1,S_2)$ between two many-body states  
being of a (Fermi) Bose type. The corresponding (anti)commutators for the Hubbard operators are
\begin{equation}
 \label{acommHub}
 \left[\hat X_{S_1S_2};\hat X_{S_3S_4}^\dagger\right]_\pm =
 \delta_{S_2,S_4}\hat X_{S_1S_3} \pm \delta_{S_1,S_3}\hat X_{S_4S_2}
\end{equation} 
where the plus sign is taken if both $\hat X_{S_1 S_2}$ and $\hat X_{S_3 S_4}$ are 
fermionic Hubbard Operators and the minus sign otherwise.
A crucial difference is appearance of an operator (contrary to number in quasiparticle case)
in the right side of (\ref{acommHub}). Nevertheless a variant of the Wick's theorem 
was developed for lattice models at equilibrium and corresponding diagrammatic techniques
were formulated and applied to studies of strongly correlated magnetically ordered 
systems~\cite{IzyumovSkryabin_1988,OvchinnikovValkov_2004}.
The goal of this paper is to adapt the methodology to nonequilibrium realm of current
carrying molecular junctions.

We consider generic model of a junction consisting of a molecule $M$ coupled to two
contacts $L$ and $R$ each characterized by its own electrochemical potential $\mu_{L,R}$
and temperature $T_{L,R}$
\begin{equation}
 \hat H = \hat H_M + \sum_{K=L,R}\left(\hat H_K + \hat V_{MK}\right)
\end{equation} 
Here $\hat H_M$ is Hamiltonian of an isolated molecule which we represent in the basis
of many-body states $\{\lvert S\rangle\}$ of the the molecule. 
$\hat H_K$ describes reservoir of free electrons and $\hat V_{MK}$ couples the two subsystems.
Explicit expressions are
\begin{align}
\label{HM}
 \hat H_M =& \sum_S E_S \hat X_{SS}
 \\
 \label{HK}
 \hat H_K =& \sum_{k\in K} \varepsilon_k\hat c_k^\dagger\hat c_k
 \\
 \label{VMK}
 \hat V_{MK} =& \sum_{k\in K}\sum_{m\in M} \left(V_{km}\hat c_k^\dagger\hat X_m + H.c.\right),
\end{align}
where $E_S$ are energies of the many-body states, $\hat X_{SS}$ are states projection
operators, $\hat c_k^\dagger$ ($\hat c_k$) is second quantization creation (annihilation)
operator for an electron in level $k$ of contacts, and $m$ are single electron (Fermi type) 
transitions between pairs of molecular states. Below we consider expansion
in the molecule-contact(s) coupling $\hat V=\sum_{K=L,R}\hat V_{MK}$.

We consider the Hubbard Green function defined on the Keldysh contour, Eq.~(\ref{HubGF}).
A multi-time correlation function can be treated similarly.
Following the usual procedure we transfer to the interaction representation with respect to 
$\hat H_0=\hat H_M+\sum_{K=L,R}\hat H_K$ and expand the scattering operator
$\hat S=\exp[-i\int_c d\tau\,\hat V(\tau)]$ (here and below operators are in the interaction picture)
in Taylor series. This leads to
\begin{align}
 \label{expand}
 & G_{(S_1S_2),(S_3S_4)}(\tau,\tau') = \sum_{n=0}^\infty\frac{(-i)^{n+1}}{n!}
 \int_cd\tau_1\ldots \int_cd\tau_n
 \nonumber \\ & \qquad\qquad
 \langle T_c\, \hat X_{S_1S_2}(\tau)\,\hat X_{S_3S_4}^\dagger(\tau')\,
 \hat V(\tau_1)\ldots\hat V(\tau_n)\rangle_0
\end{align}
Subscript $0$ indicates that evolution in each term of the expansion is governed by the 
Hamiltonian $\hat H_0$ representing uncoupled molecule and contacts. 
In the usual spirit of NEGF technique we assume that molecule and contacts (originally decoupled)
are connected at time $t\to-\infty$, and that original (before coupling) state of the system
is a direct product of states of its parts, 
$\hat\rho_0=\hat \rho^L_0\otimes\hat\rho^M_0\otimes\hat\rho^R_0$.
Thus quantum mechanical and statistical average, 
$\langle\ldots\rangle_0\equiv\mbox{Tr}[\ldots\hat\rho_0]$, in (\ref{expand})
can be splitted into the product of averages of molecular Hubbard operators and
contact quasiparticle operators. The latter satisfy the usual Wick's theorem~\cite{FetterWalecka_1971} 
and thus can be represented as the sum of all possible products of pair correlation functions 
thus yielding products of molecular self-energies due to coupling to the contacts
\begin{equation}
 \label{SE}
 \sigma^K_{mm'}(\tau,\tau') = \sum_{k\in K} V_{mk}\, g_k(\tau,\tau')\, V_{km'}
\end{equation}
where $g_k(\tau,\tau')\equiv-i\langle T_c\,\hat c_k(\tau)\,\hat c_k^\dagger(\tau')\rangle$
is NEGF of free electron in level $k$ of contact $K$.

To evaluate the average of Hubbard operators we make two additional assumptions:
1.~originally (at $t\to-\infty$) molecule was in thermal equilibrium and 
2.~after coupling to the contacts the system reached steady-state, i.e. memory of initial state was lost.
The latter is a usual assumption within the NEGF, and thus the former is unimportant for 
long time behavior of the system. Choice of thermal equilibrium as initial condition for 
the molecule allows employing diagrammatic technique of Refs.~\cite{IzyumovSkryabin_1988,OvchinnikovValkov_2004}
for evaluation of the average of Hubbard (molecular) operators.
Because the (anti)commutator of two Hubbard operators is an operator, Eq.~(\ref{acommHub}),
the Wick's theorem for Hubbard operators differs from its standard quasiparticle analog.
First one distinguishes between {\em non-diagonal} (e.g., transition $(S_1,S_2)$ with states 
$\lvert S_1\rangle$ and $\lvert S_2\rangle$ different by a number of electrons) and
{\em diagonal} Hubbard operators. 
Then pair contractions in a multi-time correlation function are created starting from 
one chosen non-diagonal operator (usually of annihilation type, i.e. such that number of electrons 
in state $\lvert S_1\rangle$ is less than the number in $\lvert S_2\rangle$) 
and continue sequentially following a set of rules regarding
choice of a non-diagonal operator to start contraction at every step of the procedure.
It is important to note that diagrammatic rules for contraction of Hubbard operators are not unique,
that is the same correlation function can be formally written in several different 
(formally equivalent) ways depending on choice of {\em a system of priorities} for the operators~\cite{IzyumovSkryabin_1988}.
Another approach utilizes {\em a principle of topological continuity}: each next contraction 
starts from the operator obtained in the previous contraction step, if the operator is
non-diagonal; otherwise contraction starts from a non-diagonal operator connected to the
already contracted part by interaction line~\cite{OvchinnikovValkov_2004}.
The latter choice allows to formulate the most general form of diagrammatic rules, however
resulting diagrams are hard to analyze in terms of which physical processes they represent.
In our analysis we will use a mixture of the two traditional choices.

Next we introduce the contraction rules for the diagrammatic technique
\begin{enumerate}
\item\label{s1} Contraction starts from the entrance point.
In the Hubbard Green function (\ref{HubGF}) operator $\hat X_{S_1S_2}(\tau)$
is the entrance point, while operator $\hat X_{S_3S_4}^\dagger(\tau')$
is the exit point. For a multi-time correlation function one can have several 
entrance (annihilation operator) and exit (creation operator) points; 
in this case contractions can start from any entrance point.
\item\label{s2} An annihilation operator resulting from $k$th contraction is the starting point
of $(k+1)$th contraction.
\item\label{s3} If result of $k$th contraction is not an annihilation operator, then we chose an
annihilation operator connected to the contracted part by interaction line as the starting point
of $(k+1)$th contraction. Note that in our case interaction is self-energy due to coupling
to contacts, Eq.~(\ref{SE}).
\item\label{s4} If neither \ref{s2}, nor \ref{s3} is possible, then $(k+1)$th contraction starts from one of 
unconnected annihilation operators chosen in accordance with accepted system
of priorities (see Section~\ref{qdmodel} for details).  
\item Single contraction from operator $\hat X_{m_1}$ to operator $\hat X_{m_2}$ 
(where $m_i$ is an index for a pair of states) is
\begin{align}
&\contraction{ \langle T_c\ldots}{\hat X }{_{m_2}(\tau_2)\dots}{\hat X}
 \langle T_c\ldots \hat X_{m_2}(\tau_2) \ldots \hat X_{m_1}(\tau_1) \ldots\rangle_0 =
 \\ & \quad
 (-1)^P i g^{(0)}_{m_1}(\tau_1,\tau_2)\langle\ldots [\hat X_{m_1};\hat X_{m_2}]_\pm(\tau_2)\ldots\rangle_0
 \nonumber
\end{align}
Here $g^{(0)}_{m_1}(\tau_1,\tau_2)$ is zero order (in the absence of molecule-contacts coupling)
Green function and $[\hat X_{m_1};\hat X_{m_2}]_\pm$ is operator of spectral weight
defined in Eq.~(\ref{acommHub}).
$P$ depends on the type of the operator $\hat X_{m_1}$:
for Bose type $P=0$; for Fermi type $P$ is the number of permutations of $\hat X_{m_1}$ with 
other Fermi operators in the correlation function of transition from original place of $\hat X_{m_1}$
to position in front of $\hat X_{m_2}$.
Note that zero order Green function of Hubbard operators $G^{(0)}_{m_1}(\tau_1,\tau_2)$
differs from $g^{(0)}_{m_1}(\tau_1,\tau_2)$ by the spectral weight factor:
$G^{(0)}_{m_1}(\tau_1,\tau_2)=g^{(0)}_{m_1}(\tau_1,\tau_2)\langle [\hat X_{m_1};\hat X_{m_1}]_\pm\rangle_0$ (the latter is called Green function~\cite{IzyumovSkryabin_1988}, 
propagator~\cite{OvchinnikovValkov_2004}, or locator~\cite{Fransson_2010} in the literature).
\item The sequence of contractions continues until all operators remaining in the correlation function
are diagonal. Separate expansion should be constructed for these correlation functions.
\item Only connected diagrams are retained in the expansion. 
\item After expansion to desired order of the coupling is finished the diagrams are
dressed in complete analogy with the standard diagrammatic techniques.
\end{enumerate}

In the resulting diagrams one can distinguish three types of contributions:
\begin{enumerate}
\item Self-energy contributions $\Sigma(\tau,\tau')$ 
- irreducible (those which cannot be cut by one Green function line) 
parts of diagrams connected to the entrance and exit points,  
respectively operators $\hat X_{S_1S_2}(\tau)$ and $\hat X_{S_3S_4}^\dagger(\tau')$ 
in Eq.~(\ref{HubGF}), by Green function lines. 
\item Spectral weight contributions $F(\tau)$ 
- parts of diagrams where entrance and exit points are connected by single Green function line.
\item Vertex contributions $\Delta(\tau,\tau')$
- parts of diagrams connected to the entrance (but not exit) point by single Green function line.
\end{enumerate}
The sum of the spectral weight and the vertex diagrams, 
$P(\tau,\tau')=\delta(\tau,\tau')F(\tau)+\Delta(\tau,\tau')$, 
is called the strength operator~\cite{OvchinnikovValkov_2004}. 

\begin{figure}[t]
\centering\includegraphics[width=\linewidth]{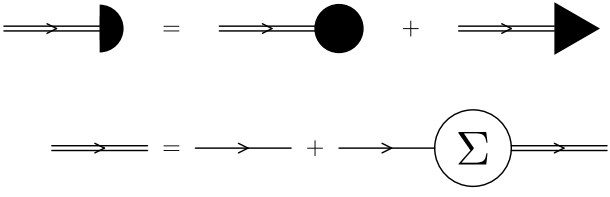}
\caption{\label{fig2}
Graphical representation of the modified Dyson equation, Eqs.~(\ref{Dyson1})-(\ref{Dyson2}).
Directed single solid line represents zero order Green function $g^{(0)}$,
directed double line stands for the dressed Green function $g$,
spectral weight $F$ is indicated with the circle, and vertex $\Delta$ with triangle.
Semi-circle represents the strength operator $P$.
}
\end{figure}

Expansion series can be resumed into a modified Dyson type equation
\begin{align}
\label{Dyson1}
G_{mm'}(\tau,\tau') =& \sum_{m_1}\int_c d\tau_1\, g_{mm_1}(\tau,\tau_1)\, P_{m_1m'}(\tau_1,\tau')
\\
\label{Dyson2}
g_{mm'}(\tau,\tau') =& g^{(0)}_{mm'}(\tau,\tau') 
+ \sum_{m_1,m_2}\int_c d\tau_1\int_c d\tau_2\, 
\\ &
g^{(0)}_{mm_1}(\tau,\tau_1)\,\Sigma_{m_1m_2}(\tau_1,\tau_2)\,g_{m_2m'}(\tau_2,\tau')
\nonumber
\end{align}
Also here indices $m$ indicate single-electron transitions between many-body states of
the system. 
Graphical representation of the equations is shown in Fig.~\ref{fig2}. 
Like the Dyson equation of the standard NEGF, Eqs.~(\ref{Dyson1})-(\ref{Dyson2})
are exact; their form reflects resummation of all the diagrams in perturbative expansion.
Similar to the Dyson equation, where the approximation enters via
a particular form of the self-energy, Eqs.~(\ref{Dyson1})-(\ref{Dyson2}) become approximate
due to particular forms of the self-energy $\Sigma$ and strength operator $P$.

Note that similar structure of Green function (without vertex contribution) was
obtained within equation-of-motion approach for retarded projection of the usual 
Green function, Eq.~(\ref{GFNEGF}),  in Ref.~\cite{HaugJauho_2008};
nonequilibrium version was derived in Ref.~\cite{GalperinNitzanRatnerPRB07}. 
Note also that Eqs.~(\ref{Dyson1})-(\ref{Dyson2}) are similar to those obtained
within auxiliary fields approach~\cite{Fransson_2010}.
However, contrary to the auxiliary fields approach to Hubbard NEGF diagrammatic formulation
yields a clear procedure of perturbative accounting for the system-bath coupling,
which is applicable also to multi-time correlation functions. In particular, already 
at second order in the coupling the diagrammatic technique yields vertex contribution,
which is shown below to be  crucial for both accuracy of the results and
the very ability to predict co- and pair-tunneling in junctions.


\section{Quantum dot model}\label{qdmodel}
We now specify to a quantum dot junction. Molecular (quantum dot) subspace is spanned
by four many-body states states $\lvert S\rangle$ ($S\in\{0,a,b,2\}$) of the following second 
quantized form
\begin{equation}
 \lvert 0\rangle\equiv \lvert 0,0\rangle \quad \lvert a\rangle\equiv\lvert 1,0\rangle\quad 
 \lvert b\rangle\equiv \lvert 0,1\rangle \quad \lvert 2\rangle\equiv\lvert 1,1\rangle
\end{equation}
Energies of the states are
$E_0=0$, $E_a=\varepsilon_a$, $E_b=\varepsilon_b$, and $E_2=\varepsilon_a+\varepsilon_b+U$,
respectively.
There are four Fermi type transitions $m\in\{1,2,3,4\}$ in the model corresponding to charge 
transfer between molecule and contacts
\begin{equation}
 \label{transitions}
 1.~\lvert 0\rangle\langle a\rvert\quad 2.~\lvert b\rangle\langle 2\rvert\quad
 3.~\lvert 0\rangle\langle b\rvert\quad 4.~\lvert a\rangle\langle 2\rvert
\end{equation}
 
\begin{figure}[t]
\centering\includegraphics[width=\linewidth]{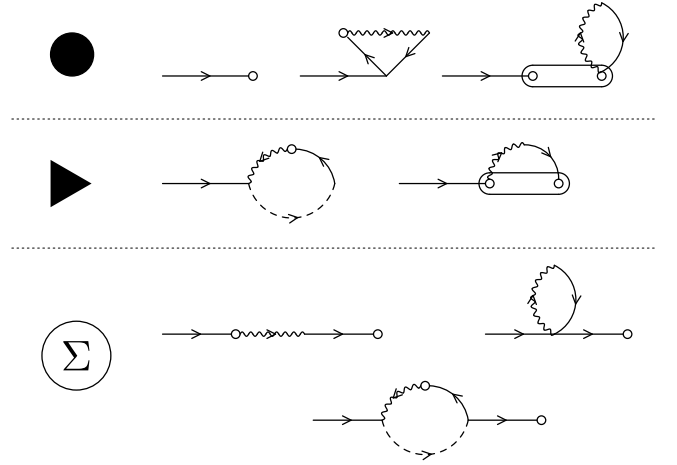}
\caption{\label{fig3}
Non-dressed diagrams up to second order in molecule-contacts coupling
for Fermi Green function $G_{mm'}$, Eq.~(\ref{Dyson1}),
contributing to the spectral weight $F$ (circle, top panel), vertex $\Delta$ (triangle, middle panel),
and self-energy $\Sigma$ (bottom panel).
Solid line represents Fermi type Green function $g^{(0)}_m$, wavy line is the interaction (\ref{SE}),
dashed line represents Bose type Green function for two-particle scattering $d^{(0)}_{02}$,
and oval stands for the correlation function $C^{(0)}$. 
}
\end{figure}

Performing expansion up to second order in molecular coupling to contacts for 
the Hubbard Green function (\ref{HubGF}) and evaluating contractions following the rules 
discussed above leads to the set of diagrams presented in Fig.~\ref{fig3}.
Diagrams for the spectral weight $F_{mm'}(\tau)$ (top panel in Fig.~\ref{fig3}) 
can be disregarded, because the latter is related to lesser and greater 
projections of the Hubbard Green functions 
\begin{equation}
 \label{F}
 F_{mm'}(t) \equiv \langle\{\hat X_m(t);\hat X_{m'}^\dagger(t)\}\rangle
 = iG_{mm'}^>(t,t)-iG_{mm'}^{<}(t,t)
\end{equation}
Dressed versions of contributions to the vertex and self-energy (middle and bottom
panels of Fig.~\ref{fig3}, respectively) are
\begin{align}
\label{Delta2}
 &\Delta_{mm'}(\tau,\tau') = \sum_{m_1,m_2}\int_c d\tau_1
 \\ & \quad
 \bigg(i(-1)^{m'} d_{02}(\tau,\tau')\, g_{\bar m' m_2}(\tau',\tau_1)\,
 F_{m_2m_1}(t_1)\,\sigma_{m_1\bar m}(\tau_1,\tau)
 \nonumber \\ & \quad
 +C_{mm_1,m_2m'}(\tau,\tau')\,\sigma_{m_1m3}(\tau,\tau_1)\, g_{m_3m_2}(\tau_1,\tau')
 \bigg)
 \nonumber 
 \end{align}
 \begin{align}
 \label{Sigma2}
& \Sigma_{mm'}(\tau,\tau') = 
 \sum_{m_1} F_{mm_1}(t)\,\sigma_{m_1m'}(\tau,\tau')
 \\ & \quad
 +i\delta(\tau,\tau')\sum_{m_1,m_2,m_3}\eta(m_3,mm_1)\,\delta_{\gamma(m_3,mm_1),m'}
\nonumber \\ &\qquad\qquad\qquad \times
 \int_c d\tau_1\,\sigma_{m_1m_2}(\tau,\tau_1)\, G_{m_2m_3}(\tau_1,\tau)
\nonumber \\ & \quad
 +i(-1)^{m'} d_{02}(\tau,\tau')\,\sigma_{\bar m'\bar m}(\tau',\tau)
 \nonumber
\end{align}
Here $\bar m=3-m$, 
$\eta(m,m_1m_2)$ and $\gamma(m,m_1m_2)$ are defined in (\ref{etadef}) and (\ref{gammadef}),
$d_{02}(\tau,\tau')$ is Green function (locator) corresponding
to two-electron Hubbard Green function
\begin{equation}
\label{D02GF}
D_{02}(\tau,\tau')=-i\langle T_c\,\hat X_{02}(\tau)\,\hat X_{02}^\dagger(\tau')\rangle,
\end{equation}
and
\begin{align}
\label{CF}
C_{m_1m_2,m_3m_4}(\tau,\tau')=\langle T_c\, \delta \hat F_{m_1m_2}(\tau)\, \delta \hat F_{m_3m_4}(\tau')\rangle
\end{align}
($\delta\hat F\equiv\hat F- \langle \hat F\rangle$) is the correlation function.
In derivation of Eqs.~(\ref{Delta2})-(\ref{Sigma2}) we utilized commutation relations 
presented in Appendix~\ref{appA}. 
Fourth order expressions are written in Appendix~\ref{appAB}.
Similarly, diagrammatic expansions are performed for the functions 
(\ref{D02GF}) and (\ref{CF}). We discuss them in Appendices~\ref{appB} and \ref{appC},
respectively.

Comparing with previous works on the Hubbard NEGF~\cite{MGNitzanRatnerPRB08,Fransson_2010,FranssonPRB02,FranssonPhotonNanostruct04,FranssonPRB05,SandalovJPCM06,SandalovPRB07} we note that considerations there were 
restricted to the first and (sometimes also) second diagram in the bottom panel
of Fig.~\ref{fig3} - correspondingly first Hubbard (HIA) and  one-loop approximations. 
Below we show that other second order diagrams are also important.
In particular, second diagram in the middle panel of Fig.~\ref{fig3} is crucial in obtaining
better approximations of exact results in the single electron tunneling regime 
and is an inherent part of description of cotunneling in junctions, 
while first diagram in the middle and last diagram in the bottom
panels of Fig.~\ref{fig3} are responsible for pair electron tunneling.
We also note that contrary to a common perception~\cite{PedersenPRB09} 
HIA is not a lowest order approximation of the nonequilibrium atomic limit.


\begin{figure}[htpb]
\centering\includegraphics[width=\linewidth]{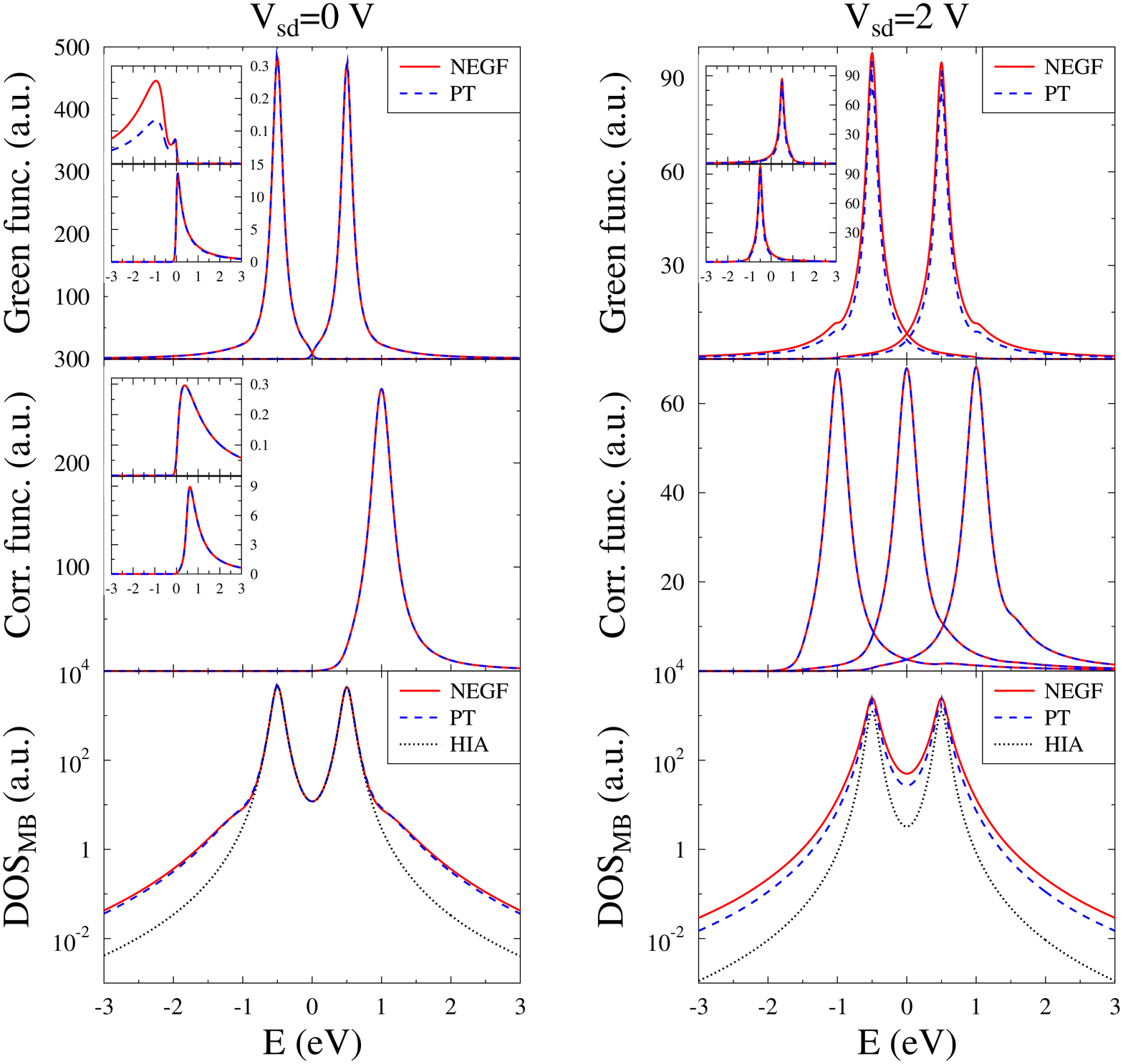}
\caption{\label{fig4}
Non-degenerate, $\varepsilon_a\neq\varepsilon_b$, two-level system (HOMO-LUMO) junction.
Exact NEGF results (solid line, red) are compared with diagrammatic perturbation theory (PT)
simulations (dashed line, blue) for equilibrium ($V_{sd}=0$, left) and nonequilibrium 
($V_{sd}=2$~V, right) junctions. 
Top graphs show Green functions $\mbox{Im}\,G^{<}_{11}(E)$, $-\mbox{Im}\,G^{>}_{44}(E)$
(main panel), $\mbox{Im}\,G^{<}_{33}(E)$ (top inset), and $-\mbox{Im}\,G^{>}_{11}(E)$ (bottom inset).
Middle graphs show correlation functions $C^{>}_{31,13}(E)$ (rightmost peak in the main panel),
$C^{>}_{13,31}(E)$ (top inset or leftmost peak in the main panel), and
$C^{>}_{33,33}(E)$ (bottom inset or central peak in the main panel).
Bottom graphs show many-body spectral function, $i\sum_m\left(G^{>}_{mm}(E)-G^{<}_{mm}(E)\right)$;
dotted line (black) shows results of HIA calculations. See text for parameters.
}
\end{figure}

\section{Numerical results}\label{numres}
Here we present simulations within the generic quantum dot model of Section~\ref{qdmodel}  
illustrating viability of the diagrammatic perturbation technique. 
We start from a non-interacting case, $U=0$, where exact solution is known
from the usual NEGF. Simulations within the perturbation theory (PT) will be compared
to the exact results.

Figure~\ref{fig4} shows results for a non-degenerate two-level system junction.
Parameters of the simulations are $T=300$~K, $\varepsilon_a=-0.5$~eV,
$\varepsilon_b=0.5$~eV, $U=0$, $\Gamma^K_{aa}=\Gamma^K_{bb}=0.1$~eV and
$\Gamma^K_{ab}=\Gamma^K_{ba}=0$ ($K=L,R$). Fermi energy is taken as origin,
$E_F=0$, and bias is applied symmetrically $\mu_{L/R}=E_F\pm V_{sd}/2$.
Calculations are performed on a grid spanning region from $-6$ to $6$~eV
with step $0.005$~eV. One sees that diagrammatic perturbation theory 
(dashed lines in Fig.~\ref{fig4}) reproduces exact results (solid lines in Fig.~\ref{fig4}) 
quite accurately. Taking into account that we build our consideration on top of
the equilibrium theory~\cite{IzyumovSkryabin_1988,OvchinnikovValkov_2004}, 
it is interesting to note that in some cases nonequilibrum results
appear to be closer to exact data (compare top insets in the left and right top panels of 
Fig.~\ref{fig4}). Another important observation comes from comparison of the PT
and HIA results to exact data (compare dashed and dotted to solid lines in bottom panels 
of Fig.~\ref{fig4}). In this calculation (with dominance of single electron tunneling) 
the main difference between the two approximations comes from second diagram
in the middle panel of Fig.~\ref{fig3}. This diagram was omitted in most previous
Hubbard NEGF considerations. 

\begin{figure}[htpb]
\centering\includegraphics[width=\linewidth]{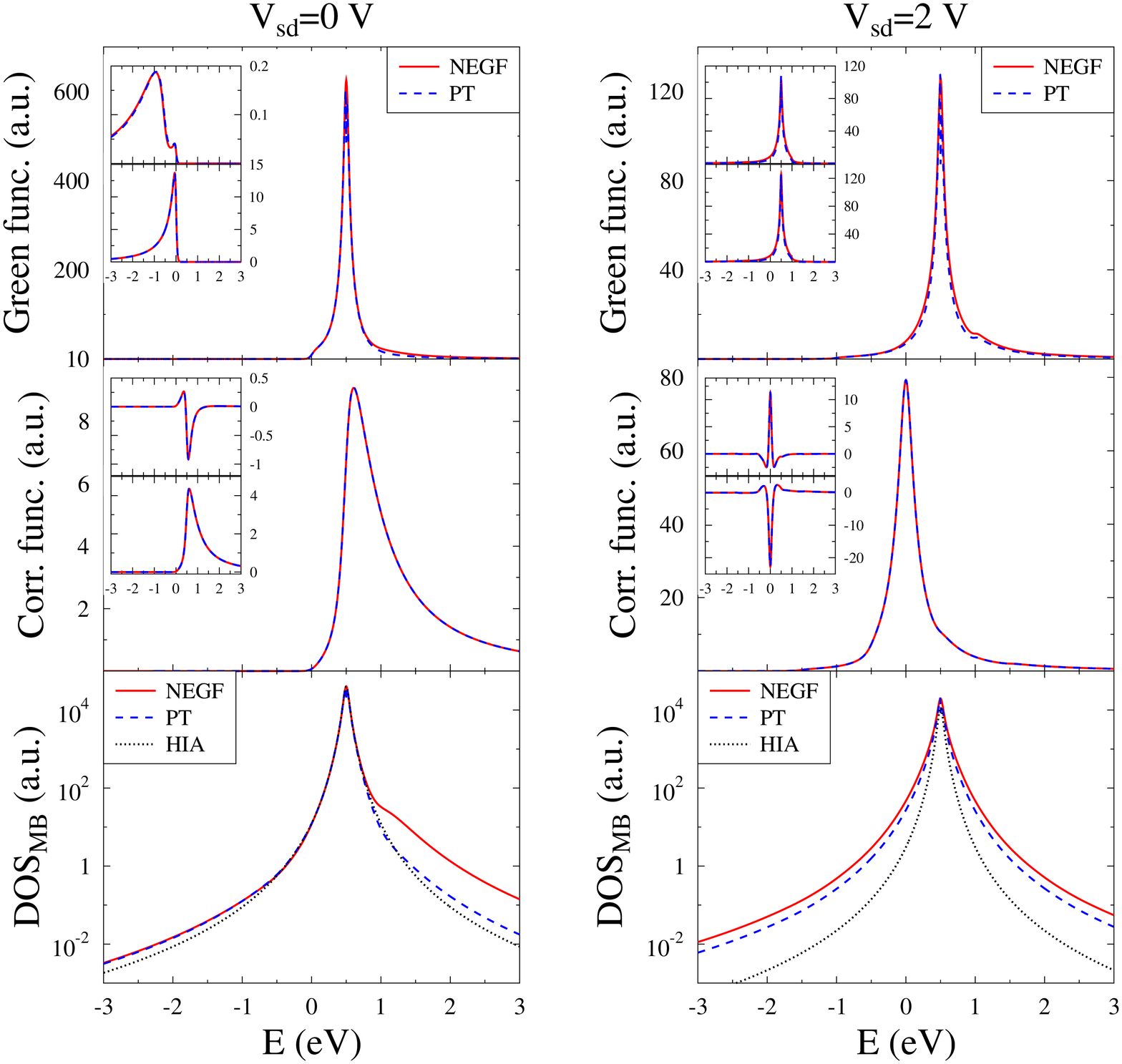}
\caption{\label{fig5}
Degenerate, $\varepsilon_a=\varepsilon_b$, two-level system (HOMO-LUMO) junction.
Exact NEGF results (solid line, red) are compared with diagrammatic perturbation theory (PT)
simulations (dashed line, blue) for equilibrium ($V_{sd}=0$, left) and nonequilibrium 
($V_{sd}=2$~V, right) junctions. 
Top graphs show Green functions $-\mbox{Im}\,G^{>}_{11}(E)$ (main panel), 
$\mbox{Im}\,G^{<}_{22}(E)$ (top inset), and $\mbox{Im}\,G^{<}_{11}(E)$ (bottom inset).
Middle graphs show correlation functions $C^{>}_{33,33}(E)$ (main panel),
$C^{>}_{13,13}(E)$ (top inset), and $C^{>}_{13,44}(E)$ (bottom inset).
Bottom graphs show many-body spectral function, $i\sum_m\left(G^{>}_{mm}(E)-G^{<}_{mm}(E)\right)$;
dotted line (black) shows results of HIA calculations. See text for parameters.
}
\end{figure}
\begin{figure}[htpb]
\centering\includegraphics[width=\linewidth]{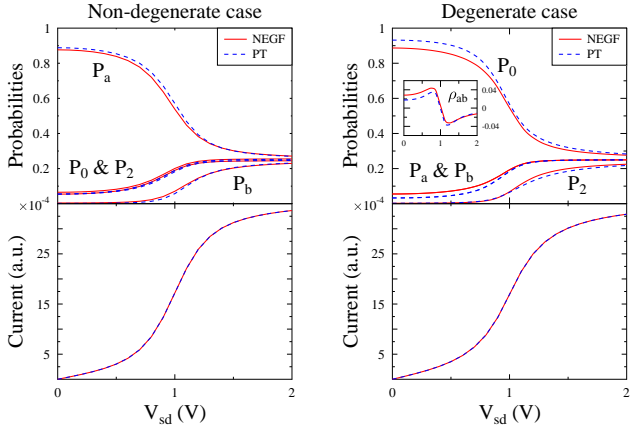}
\caption{\label{fig6}
Non-interacting junction. 
Exact NEGF results (solid line, red) are compared with diagrammatic perturbation theory (PT)
simulations (dashed line, blue).
Shown are state probabilities (top panels) and current (bottom panels) vs. applied bias $V_{sd}$
for non-degenerate (left) and degenerate (right) cases of Figs.~\ref{fig4} and \ref{fig5}, respectively.
Inset in the top right panel shows coherence vs. applied bias. See text for parameters.
}
\end{figure}

Results for degenerate two-level system are presented in Fig.~\ref{fig5}.
Here $\varepsilon_a=\varepsilon_b=0.5$~eV and $\Gamma^K_{ab}=\Gamma^K_{ba}=0.05$~eV
($K=L,R$). Other parameters are as in Fig.~\ref{fig4}. Also here PT theory demonstrates 
good approximation to the exact results and an advantage as compared to the HIA simulations.

Figure~\ref{fig6} shows currents and state probabilities as functions of applied bias for the two 
cases presented in Figs.~\ref{fig4}-\ref{fig5}. One sees that PT yields reasonable approximation
to the exact results including the coherence (see inset in the right top panel) and perfect agreement 
for current-voltage characteristic.

\begin{figure}[t]
\centering\includegraphics[width=0.8\linewidth]{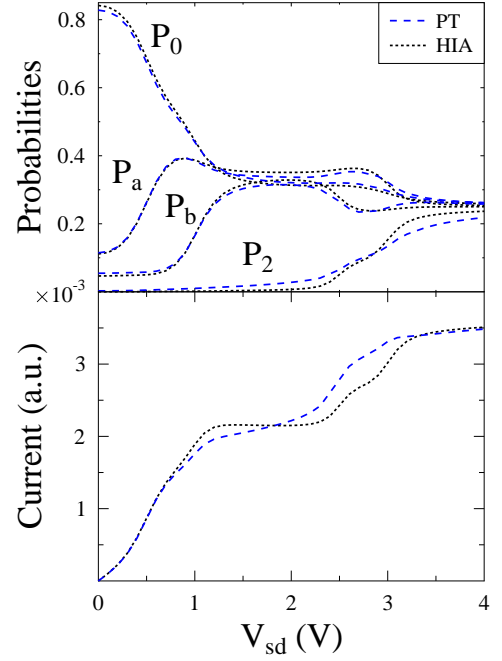}
\caption{\label{fig7}
Quantum dot junction in the Coulomb blockade regime 
Diagrammatic perturbation theory (PT) simulations (dashed line, blue) are compared with 
the HIA (dotted line, black).
Shown are state probabilities (top panel) and current (bottom panel) vs. applied bias $V_{sd}$.
See text for parameters.
}
\end{figure}

We now consider interacting junctions, $U\neq 0$. The Hubbard NEGF treatment 
(auxiliary fields formulation) of quantum-dot junction in the Coulomb blockade regime 
was discussed in details in Ref.~\cite{FranssonPRB05}.
Figure~\ref{fig7} shows that the second order PT (Fig.~\ref{fig3}) yields similar results in this regime.
Here parameters of the simulation are  $\varepsilon_a=0.25$~eV, $\varepsilon_b=0.5$~eV,
and $U=1$~eV. Other parameters are as in Fig.~\ref{fig4}.

\begin{figure}[htpb]
\centering\includegraphics[width=0.8\linewidth]{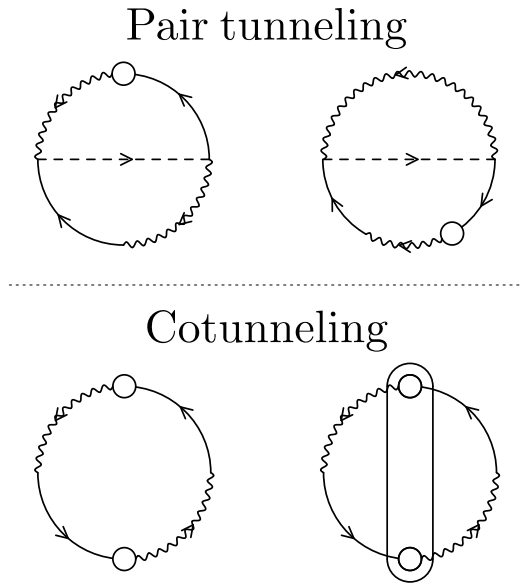}
\caption{\label{fig8}
Diagrams responsible for pair (top) and cotunneling (bottom) transport in 
the negative-U model.
}
\end{figure}

\begin{figure}[htpb]
\centering\includegraphics[width=0.8\linewidth]{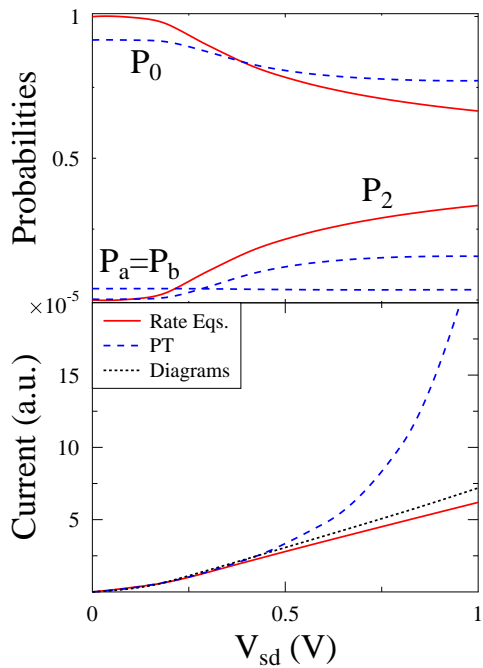}
\caption{\label{fig9}
Pair electron tunneling in junction. 
Shown are state probabilities (top panel) and current (bottom panel) vs. applied bias $V_{sd}$.
Diagrammatic perturbation theory (PT) simulations (dashed line, blue) are compared with 
the rate equation results of Ref.~\cite{KochRaikhVonOppenPRL06} (solid line, red).
Bottom panel  also presents contribution to the current from diagrams of Fig.~\ref{fig8} (dotted line, black).
See text for parameters.
}
\end{figure}

Finally, we consider regime of pair electron tunneling in junctions. Within negative-U model
this regime was examined  in Ref.~\cite{KochRaikhVonOppenPRL06} using
the Schrieffer-Wolf transformation of the Hubbard Hamiltonian and evaluating
Golden rule rates after the transformation. Noting that result of 
the Schrieffer-Wolf transformation can be effectively achieved within second order
perturbation theory~\cite{MartinMozyrskyPRB05}, and keeping in mind that  
in the Hubbard NEGF expression for current, 
\begin{align}
\label{IK}
 & I_K(t) = 2\,\mbox{Re}\,\int_{-\infty}^t dt'\,
 \\ & \qquad
 \mbox{Tr}\bigg[
 G^{>}(t,t')\,\sigma_K^{<}(t',t)-G^{<}(t,t')\,\sigma_K^{>}(t',t)
 \bigg]
 \nonumber
\end{align}
(here $K=L,R$ and $\mbox{Tr}[\ldots]$ is trace over single-electron transitions (\ref{transitions})),
two orders in coupling (\ref{VMK}) enter via self-energy $\Sigma_K$,
It is natural to expect that pair tunneling is given by second-order diagrams
in Fig.~\ref{fig3} augmented by self-energy (wavy) line of Eq.~(\ref{IK}). 
Indeed, one can show that results of Ref.~\cite{KochRaikhVonOppenPRL06} 
can be obtained from (non-dressed) diagrams presented in Fig.~\ref{fig8}
(see Appendix~\ref{appD} for details). One sees that the pair tunneling 
results from second order diagrams involving two-electron propagator
(first diagram in the middle panel and last diagram in the bottom panel of Fig.~\ref{fig3}),
while cotunneling results from the diagrams involving single-electron
propagators (last diagram of the middle panel and  first diagram
of the bottom panel of Fig.~\ref{fig3}). Note that for either pair or cotunneling only the sum of 
both self-energy and vertex diagrams is capable to provide correct results.

 Figure~\ref{fig9} compares results of the PT simulation (dashed line) with rate equation results
 of Ref.~\cite{KochRaikhVonOppenPRL06} (solid line). Calculations are performed for the negative-U 
 model with parameters $\varepsilon_a=\varepsilon_b=2$~eV and $U=-3.8$~eV.
 Other parameters are as in Fig.~\ref{fig4}. 
 We note that contrary to previous results fourth order PT is required here.
 The PT takes broadening into account which leads to small non-zero values for 
 the probabilities $P_a$ and $P_b$ of single-electron states (see top panel).
 Current-voltage characteristic (bottom panel) 
 is almost identical to that of rate equations at low bias, $V_{sd}<0.5$~V,
 and deviates at higher biases, where contribution of the single-particle tunneling
 is non-negligible. 
 

\section{Conclusion}\label{conclude}
We present a nonequilibrium flavor of diagrammatic technique for Hubbard Green functions.
The technique is suitable for description of nonequilibrium steady-states in junctions.
We assume that initial state of (uncoupled) system and baths does not impact nonequilibrium 
steady-state, which allows to utilize equilibrium considerations for the Hubbard lattice models
for evaluation of zero-order (uncoupled) correlation functions of Hubbard operators.
The latter leads to a nonequilibrium diagrammatic expansion for the Hubbard NEGF
on the Keldysh contour. Similarly, one can consider diagrammatic expansion for
 multi-time correlation functions on the contour.
 
 We illustrate viability of the approach with numerical examples of transport in 
 non-interacting (two-level system) and interacting (quantum dot)  junctions.
 For non-interacting system we compare the diagrammatic PT to exact NEGF results,
 and show that the approach is quite accurate (for both non-degenerate and degenerate
 cases) already at second order of the PT.  
Interacting calculations are compared with similar considerations available in the literature.
In particular, quantum dot junction results are compared to the auxiliary-field
approach to the Hubbard NEGF. Results of the PT simulations within the negative-U model 
for pair electron tunneling are compared with rate equations applied to Schrieffer-Wolf transformed Hamiltonian. We show importance of the correlation (vertex) diagrams, which were omitted 
in previous Hubbard NEGF considerations. 
 
The diagrammatic PT for Hubbard NEGF contributes to development of tools for the nonequilibrium atomic limit, where the response of a molecular junction to external perturbations is characterized
utilizing many-body states of the isolated molecule while coupling to the contacts is treated perturbatively. 
Such an approach yields a possibility of incorporating standard quantum chemistry 
and nonlinear optical spectroscopy methods (mostly formulated for isolated molecules and utilizing
many-body states) to description of quickly developing field of nanoscale optoelectronics 
of molecular junctions.  It may also be useful in thermodynamic studies of thermoelectric and 
photovoltaic molecular devices.  
Finally we note that while approaches utilizing many-body states description were successfully
implemented in a number {\em ab initio} simulations~\cite{HettlerPRL03,YeganehNL09,CorniaNL10,WegewijsNatPhys13,WhiteTretiakNL14}, 
combination of such formulations with quasiparticle based approaches capitalizing on 
ability of the former to treat strong local interactions and scalability of the latter would be advantageous.
This is a goal of our future research.


\begin{acknowledgments}
We thank Guy Cohen and Robert van Leeuwen for very helpful discussions.
This material is based upon work supported by the US Department of Energy under DE-SC0006422
and the National Science Foundation under CHE - 1565939.
\end{acknowledgments}


\appendix
\section{Commutation relations}\label{appA}
Here we present commutation relations between Hubbard operators utilized in derivation of diagrams.
Note that the commutation relations can be formulated in a compact form utilizing a notion
of root vectors~\cite{OvchinnikovValkov_2004}. However it is hard to keep track of the physics 
represented by the diagrams constructed utilizing the notion. Thus here we 
present explicit commutation relations of operators relevant for the quantum dot model
\begin{equation}
\label{Acomm}
\begin{split}
 & \{\hat X_m;\hat X_{m'}\} = \delta_{m,\bar m'}\hat X_{02} 
 \\
 & [\hat X_m;\hat X_{02}^\dagger] = (-1)^m\hat X_{\bar m}^\dagger 
 \\
 & [\hat X_m;\hat N] = -\hat X_m
 \\
 & [\hat X_m;\hat F_{m_1m_2}] = \eta(m,m_1m_2)\hat X_{\gamma(m,m_1m_2)}
 \\
 & [\hat X_{02};\hat X_m^\dagger] = (-1)^m\hat X_{\bar m}
 \\
 & [\hat X_{02};\hat F_{m_1m_2}] = \delta_{m_1,m_2}(-1)^{m_1}\hat X_{02}
 \\
 & [\hat X_{02};\hat N] = -2\,\hat X_{02}
\end{split}
\end{equation}
with other commutators zero.
Here $\bar m=3-m$,
\begin{align}
 &\hat F_{mm'} \equiv \{\hat X_m;\hat X_{m'}^\dagger\} 
 \qquad
 \hat N \equiv [\hat X_{02};\hat X_{02}^\dagger]
 \\
 \label{etadef}
 &\eta(m,m_1m_2) = \delta_{m_1,m_2}(-1)^{m_1}-\delta_{m_1,m_2\pm 2}(-1)^m
 \\
 \label{gammadef}
 &\gamma(m,m_1m_2) = \delta_{m_1,m_2} m + \delta_{m_1,m_2\pm 2}(m\pm 2)
\end{align}


\section{Fourth order expressions for Green function (\ref{HubGF}) self-energy and vertex}\label{appAB}
Dressed fourth order contributions to self-energy and vertex of 
Green function (\ref{HubGF}) are
\begin{widetext}
\begin{align}
&\Sigma_{mm'}^{(4)}(\tau,\tau') =-\sum_{\{m\}}\int_c d\tau_1\int_c d\tau_2\,
\\ &
\bigg(
\eta(m_5,mm_1)\,\eta(m_6,m_3m_4)\delta_{\gamma(m_6,m_3m_4),m'}
\sigma_{m_1m_2}(\tau,\tau_2)\, g_{m_2m_3}(\tau_2,\tau')\,
\sigma_{m_4m_7}(\tau,\tau_1)\, g_{m_7m_5}(\tau_1,\tau)\, 
g_{\gamma(m_5,mm_1)m_6}(\tau,\tau')
\nonumber \\ &
+(-1)^m d_{02}(\tau,\tau')\,\sigma_{\bar m'm_2}(\tau',\tau_2)
\bigg[
(-1)^{m'} d_{02}(\tau_2,\tau)\,\sigma_{mm_1}(\tau,\tau_1)\, g_{m_1\bar m_2}(\tau_1,\tau_2)
+i\, g_{m_2m_1}(\tau_2,\tau_1)\, F_{m_1m_3}(\tau_1)\,\sigma_{m_3\bar m}(\tau_1,\tau)
\bigg]
\bigg)
\nonumber \\ &
\Delta^{(4)}_{mm'}(\tau,\tau') =-\sum_{\{m\}}\int_c d\tau_1\int_c d\tau_2 \int_c d\tau_3
\\ &
\bigg(
(-1)^{m'}\eta(m_4,mm_5)\,\sigma_{m_5m_3}(\tau,\tau_3)\, g_{m_3\bar m_2}(\tau_3,\tau_2)\,
g_{\gamma(m_4,mm_5)m_1}(\tau,\tau_1)\, F_{m_1m_6}(\tau_1)\,
\sigma_{m_6m_2}(\tau_1,\tau_2)\, d_{02}(\tau_2,\tau')\, g_{\bar m'm_4}(\tau',\tau)
\nonumber \\ &
+(-1)^{m+m'} d_{02}(\tau,\tau')\, g_{\bar m' m_1}(\tau',\tau_1)\, F_{m_1m_4}(\tau_1)\,
\sigma_{m_4m_2}(\tau_1,\tau_2)\, d_{02}(\tau_2,\tau)\, \sigma_{mm_3}(\tau,\tau_3)\,
g_{m_3\bar m_2}(\tau_3,\tau_2)
\nonumber \\ &
+i\, \eta(m_7,m_6,m')\, C_{mm_5,m_1m_4}(\tau,\tau_1)\,
\sigma_{m_5m_3}(\tau,\tau_3)\, g_{m_3m_6}(\tau_3,\tau')\,
g_{\gamma(m_7,m_6m')\, m_1}(\tau',\tau_1)\, 
\sigma_{m_4m_2}(\tau_1,\tau_2)\, g_{m_2m_7}(\tau_2,\tau')
\nonumber \\ &
-i\, (-1)^{m'} C_{mm_5,m_1m_4}(\tau,\tau_1)\, \sigma_{m_5m_3}(\tau,\tau_3)\,
g_{m_3\bar m_2}(\tau_3,\tau_2)\,\sigma_{m_4m_2}(\tau_1,\tau_2)\, d_{02}(\tau_2,\tau')\,
g_{\bar m' m_1}(\tau',\tau_1)
\nonumber \\ &
+i\,\eta(m_7,mm_5)\, C_{m_1m_4,m_6m'}(\tau_1,\tau')\,
\sigma_{m_5m_3}(\tau,\tau_3)\, g_{m_3m_6}(\tau_3,\tau')\,
g_{\gamma(m_7,mm_5)\, m_1}(\tau,\tau_1)\, \sigma_{m_4m_2}(\tau_1\tau_2)\,
g_{m_2m_7}(\tau_2,\tau)
\nonumber \\ &
-i\, \eta(m_6,mm_5)\, C_{m_1m_4,m_6m'}(\tau_1,\tau')\,
\sigma_{m_5m_3}(\tau,\tau_3)\, g_{m_3,m_1}(\tau_3,\tau_1)\,
\sigma_{m_4m_2}(\tau_1,\tau_2)\, g_{m_2m_7}(\tau_2,\tau)\,
g_{\gamma(m_6,mm_5)\, m_8}(\tau,\tau')
\nonumber \\ &
-i (-1)^{m_1} C_{m_4m_3.m_6m'}(\tau_3,\tau')\,
\sigma_{m_3 \bar m}(\tau_3,\tau)\, d_{02}(\tau,\tau_1)\, g_{\bar m_1 m_6}(\tau_1,\tau')\,
\sigma_{m_1m_2}(\tau_1,\tau_2)\, g_{m_2m_4}(\tau_2,\tau_3)
\nonumber \\ &
+i (-1)^{m_1} C_{m_4m_3,m_6m'}(\tau_3,\tau')\,
\sigma_{m_3\bar m}(\tau_3,\tau)\, d_{02}(\tau,\tau_1)\,
\sigma_{m-1m_2}(\tau_1,\tau_2)\, g_{m_2m_6}(\tau_2,\tau')\,
g_{\bar m_1m_4}(\tau_1,\tau_3)
\bigg)
\nonumber
\end{align}
\end{widetext}


\section{Diagrammatic expansion for Green function (\ref{D02GF})}\label{appB}
\begin{figure}[t]
\centering\includegraphics[width=0.8\linewidth]{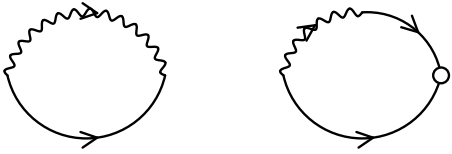}
\caption{\label{fig10}
Second order (non-dressed) contributions to self-energy $\Sigma_{02}$ (left) 
and vertex $\Delta_{02}$ (right).
}
\end{figure}

Diagrammatic expansion for the two-particle Green function follows the same rules 
as for the single-particle Green function. Second order diagrams for the self-energy
$\Sigma_{02}$ and vertex $\Delta_{02}$ are shown in Fig.~\ref{fig10}.
Explicit dressed expressions are
\begin{align}
\Sigma^{(2)}_{02}(\tau,\tau') =& i \sum_{m_1,m_2} (-1)^{m_1}\sigma_{m_1m_2}(\tau,\tau')\,
g_{\bar m_1\bar m_2}(\tau,\tau')
\\
\Delta^{(2)}_{02}(\tau,\tau') =& -i\sum_{\begin{subarray}{c}m_1,m_2\\m_3,m_4\end{subarray}}
(-1)^{m_1+m_3} g_{\bar m_1m_3}(\tau,\tau')
\\ \times&
\int_c d\tau_1\,\sigma_{m_1m_2}(\tau,\tau_1)\, g_{m_2m_4}(\tau_1,\tau')\, F_{m_4\bar m_3}(\tau')
\nonumber
\end{align} 

Dressed fourth order contributions are
\begin{align}
& S_{02}^{(4)}(\tau,\tau') =
 \sum_{\{m\}} (-1)^m\int_c d\tau_1 \int_c d\tau_2\, 
\\ &
\bigg( g_{m_1m_2}(\tau_1,\tau_2)\, g_{\bar m,\bar m'}(\tau,\tau')
- g_{m_1\bar m'}(\tau_1,\tau')\, g_{\bar mm_2}(\tau,\tau_2) 
\bigg)
\nonumber  \\ & \times
 \sigma_{m m_1}(\tau,\tau_1)\, F_{m_2m_3}(\tau_2)\,\sigma_{m_3m'}(\tau_2,\tau')
\nonumber 
\\
& \Delta_{02}^{(4)}(\tau,\tau') = i\sum_{\{m\}} (-1)^m \int_cd\tau_1\int_c d\tau_2\int_c d\tau_3\,
\\ &
\bigg( 
g_{m_1m_2}(\tau_1,\tau_2)\, g_{\bar m \bar m_4}(\tau,\tau_3)
-g_{m_1\bar m_4}(\tau_1,\tau_4)\, g_{\bar mm_2}(\tau,\tau_2)\,
\bigg)
\nonumber \\ & \times
\sigma_{mm_1}(\tau,\tau_1)\,\sigma_{m_3m_4}(\tau_2,\tau_3)\,d_{02}(\tau_3,\tau')\, C_{02\,02,m_2m_3}(\tau',\tau_2)
\nonumber
\end{align}
Note that 
\begin{equation}
C_{02\,02,m_2m_3}(\tau,\tau')\equiv C_{0S\,0S,m_2m_3}(\tau,\tau')-C_{S2\,S2,m_2m_3}(\tau,\tau')
\end{equation}
with $S=a$ or $b$.


\section{Diagrammatic expansion for correlation function (\ref{CF})}\label{appC}
Diagrammatic expansion for correlation functions follows contraction rules formulated in
Section~\ref{diag}. Resummation of the diagrams was discussed in Ref.~\cite{IzyumovKatsnelsonSkryabin_1994}.

In actual calculations we only utilized a single bubble diagram.
Its explicit expression is 
\begin{align}
& C_{m_1m_2,m_3m_4}(\tau,\tau') =
\sum_{m_5,m_6}\eta(m_5,m_1m_2)\,\eta(m_6,m_3m_4)
\\ & \qquad \qquad
g_{\gamma(m_5,m_1m_2)\, m_6}(\tau,\tau')\,
g_{\gamma(m_6,m_3m_4)\, m_5}(\tau',\tau)
\nonumber
\end{align}
Numerical results presented in Section~\ref{numres} show  that the approximation
appears to be quite accurate.


\section{Pair and cotunneling diagrams}\label{appD}
Here we prove that diagrams of Fig.~\ref{fig8} represent pair and cotunneling.

Second order contribution in molecule-contacts coupling, Eq.~(\ref{VMK}),  to
the Hubbard Green function, Eq.~(\ref{HubGF}), is
\begin{align}
\label{HubGF2}
& G_{mm'}^{(2)}(\tau,\tau')= 
\sum_{\begin{subarray}{c}m_1,k_1,\sigma_1\\m_2,k_2,\sigma_2\end{subarray}}
\int_c d\tau_1 \int_c d\tau_2\, V_{k_1\sigma_1,m_1}\, V_{m_2,k_2\sigma_2}
\\ &
\langle T_c\, \hat X_{m}(\tau)\,\hat X_{m'}^\dagger(\tau')\,
\hat c_{k_1\sigma_1}^\dagger(\tau_1)\,\hat X_{m_1}(\tau_1)\,
\hat X_{m_2}^\dagger(\tau_2)\,\hat c_{k_2\sigma_2}(\tau_2)\rangle_0
\nonumber
\end{align}
We now can apply contraction rules of Section~\ref{diag}, which leads to a set of diagrams
presented in Fig.~\ref{fig3}. These augmented with the self-energy $\sigma_K$ lines
results in a set of contributions to the current (\ref{IK}). We then project these contributions
using connection between scattering theory and the Keldysh contour formulation 
(as discussed in Ref.~\cite{MGRatnerNitzanJCP15})
and employing the Langreth (contour deformation) rules~\cite{HaugJauho_2008}.
In particular, below we show that for diagrams in Fig.~\ref{fig8} one can identify projections 
of contour variables $\tau_{1,2}$ onto real time axis $t_{1,2}$ which after substituting lesser and 
greater projections of (\ref{HubGF2}) into (\ref{IK}) yield pair tunneling and cotunneling. 

\begin{figure}[htbp]
\centering\includegraphics[width=\linewidth]{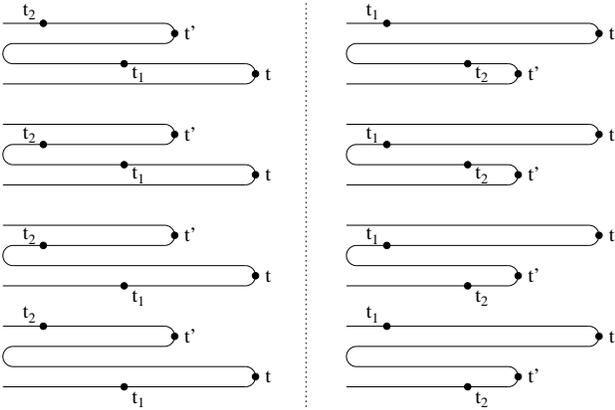}
\caption{\label{fig11}
Langreth projections of contour variables $\tau_{1,2}$ in Eq.~(\ref{HubGF2})
which yield pair tunneling contribution to the current, Eq.~(\ref{IK}).
Left (right) panel shows projections for the in-scattering (out-scattering) flux.
}
\end{figure}

For pair tunneling we use top diagrams of Fig.~\ref{fig8} and projections shown in Fig.~\ref{fig11} 
in Eq.~(\ref{IK}). The projections lead to integrals
\begin{equation}
\int_{-\infty}^{+\infty}d(t'-t)\ldots\int_{-\infty}^0 d(t_1-t)\ldots\int_{-\infty}^0 d(t_2-t')\ldots
\end{equation}
We then perform the integrations for the top diagrams of Fig.~\ref{fig8}
employing the following expressions for lesser and greater projections 
of zero-order propagators
\begin{align}
\label{gm0_gt}
 g_m^{(0)\, >}(t,t') =& -i \frac{P_{m(1)}}{P_{m(1)}+P_{m(2)}} e^{-iE_m(t-t')}
 \\
 \label{gm0_lt}
 g_m^{(0)\, <}(t,t') =& +i \frac{P_{m(2)}}{P_{m(1)}+P_{m(2)}} e^{-iE_m(t-t')}
 \\
 \label{d02_gt}
 d_{02}^{(0)\, >}(t,t') =& -i \frac{P_0}{P_0-P_2} e^{-iE_{02}(t-t')}
 \\
 \label{d02_lt}
 d_{02}^{(0)\, <}(t,t') =& -i \frac{P_2}{P_0-P_2} e^{-iE_{02}(t-t')}
\end{align}
Here $P_S$ ($S=0,a,b,2$) is the probability to find the system in state $\lvert S\rangle$,
$P_{m(1)}$ and $P_{m(2)}$  are probabilities for the first and second state of 
the transitions defined in  Eq.~(\ref{transitions}), $E_m\equiv E_{m(2)}-E_{m(1)}$, and
$E_{02}\equiv E_2-E_0$. 
Taking into account that for $\varepsilon_{a,b}\gg 0$ and $|U|\gg 1$ only 
$P_0$ and $P_2$ are non-zero we get the following contribution to the current (\ref{IK})
\begin{equation}
 I_K = W^K_{2\leftarrow 0}\, P_0 - W^K_{0\leftarrow 2}\, P_2
\end{equation}
with the rates
\begin{align}
 & W^K_{2\leftarrow 0} = \int \frac{dE}{2\pi}\,
 \bigg\lvert \frac{1}{E-\varepsilon+i\delta} - \frac{1}{E-\varepsilon-U+i\delta}\bigg\rvert^2
 \\ &\quad \times 
 \Gamma^K f_K(E_{02}-E)\sum_{K'=L,R}\Gamma^{K'} f_{K'}(E)
 \nonumber \\
 & W^K_{0\leftarrow 2} = \int \frac{dE}{2\pi}\,
 \bigg\lvert \frac{1}{E-\varepsilon+i\delta} - \frac{1}{E-\varepsilon-U+i\delta}\bigg\rvert^2
 \\ &\quad \times 
 \Gamma^K [1-f_K(E_{02}-E)]\sum_{K'=L,R}\Gamma^{K'}[1-f_{K'}(E)]
 \nonumber
\end{align}
Here $\varepsilon_a=\varepsilon_b\equiv\varepsilon$, 
$\Gamma^K_{aa}=\Gamma^K_{bb}\equiv\Gamma^K$, $\Gamma\equiv\Gamma^L+\Gamma^R$,
$\Gamma^K_{ab}=\Gamma^K_{ba}=0$, $f_K(E)$ is the Fermi-Dirac thermal distribution
in contact $K$, and $\delta\to 0+$.
These are exactly the pair tunneling rates originally derived in Ref.~\cite{KochRaikhVonOppenPRL06}.

\begin{figure}[htbp]
\centering\includegraphics[width=\linewidth]{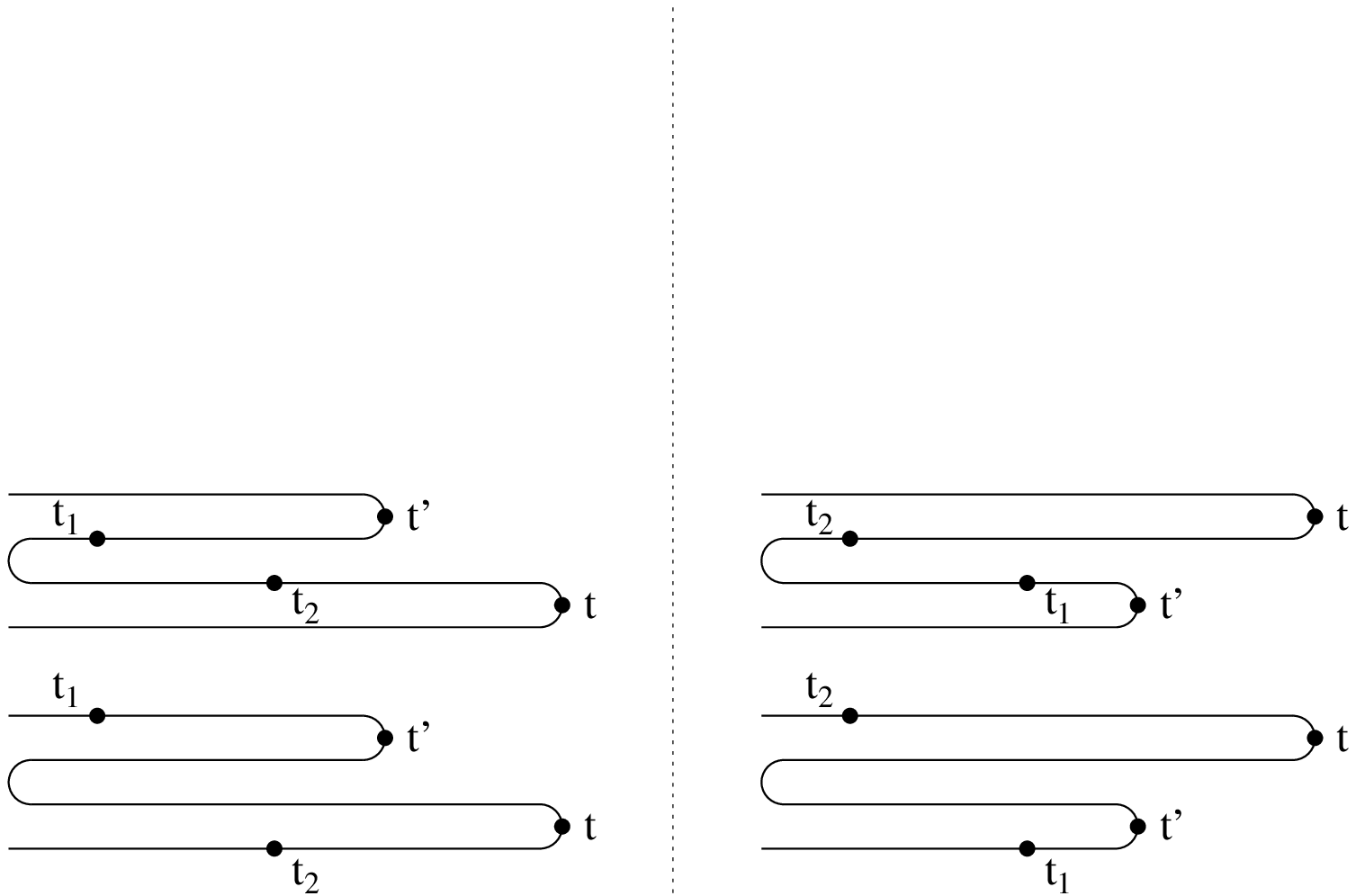}
\caption{\label{fig12}
Langreth projections of contour variables $\tau_{1,2}$ in Eq.~(\ref{HubGF2})
which yield cotunneling contribution to the current, Eq.~(\ref{IK}).
Left (right) panel shows projections for the in-scattering (out-scattering) flux.
}
\end{figure}

Similarly, employing bottom diagrams of Fig.~\ref{fig8} and projections presented in Fig.~\ref{fig12},
and utilizing Eqs.~(\ref{gm0_gt})-(\ref{d02_lt}) and zero-order expression for correlation functions
\begin{align}
& C^{(0)}_{mm,m'm'}(\tau,\tau') = \delta_{m,m'}F^{(0)}_{mm} - F_{mm}^{(0)}\, F_{m'm'}^{(0)} 
\\
& F_{mm}^{(0)} = P_{m(1)}+P_{m(2)}
\end{align}
we can derive expressions for cotunneling rates in a similar fashion. 

%

\end{document}